\documentclass[a4paper]{article}
\def\be{\begin{equation}}
\def\eea{\end{eqnarray}}
\def\bea{\begin{eqnarray}}
\def\ee{\end{equation}}
\def\d{\mathrm{d}}
\usepackage[psamsfonts]{amssymb}
\usepackage{amsmath}
\usepackage{pstricks}
\usepackage{cite}

\author{M. Alimohammadi\footnote{alimohmd@ut.ac.ir}\ \ and H.
Mohseni Sadjadi\footnote{mohsenisad@ut.ac.ir}
\\ {\small Department of Physics, University of Tehran,}
\\ {\small North Karegar Ave., Tehran, Iran.}}
\title{ The $\omega =-1$ crossing of the quintom model with
 slowly-varying potentials}
\begin{document}
\maketitle
\begin{abstract}

\noindent Considering the quintom model with arbitrary potential,
it is shown that there always exists a solution which evolves from
$\omega >-1$ region to $\omega <-1$ region. The problem is
restricted to the slowly varying potentials, i.e. the slow-roll
approximation. It is seen that the rate of this phase transition
only depends on the energy density of matter at transition time,
which itself is equal to the kinetic part of quintom energy
density at that time. The perturbative solutions of the fields are
also obtained.
\end{abstract}

\section{Introduction}
Recent astrophysical data indicate that the expansion of the
universe is accelerating \cite{acc}. This acceleration can be
related to the presence of a perfect fluid with negative pressure,
known as dark energy, which constitutes two third of our present
universe.

One candidate for dark energy is the cosmological constant: a
constant quantum vacuum energy density which fills the space
homogeneously, corresponding to a fluid with a constant equation
of state, EOS, parameter $\omega_d=-1$ (in this paper $\omega_d$
and $\omega$ denote the EOS parameters of dark energy component
and the universe, respectively).  This model suffers from fine
tuning and coincidence problem \cite{dark}. Alternatively,
dynamical homogeneous fields, e.g. scalar fields with canonical
and non-canonical kinetic terms and with various potentials, have
been proposed as the origin of dark energy \cite{quintes,phant}.
In these models, in contrast to the cosmological constant model,
the EOS parameter may vary with time. The EOS parameter satisfying
$-1<\omega<-1/3$, which determines the quintessence era, can be
achieved by introducing a normal scalar field $\phi$, known as
quintessence scalar field \cite{quintes}.

Due to the fact that the observations have shown some mild
preference for an equation of state parameter $\omega<-1$, the
phantom scalar field $\sigma$ with a wrong sign kinetic term was
introduced in the literature \cite{phant}. Depending on phantom
scalar field potential, different solutions such as asymptotic de
Sitter,  big rip, etc. may be obtained \cite{fa}.

Some astrophysical data seem to slightly favor an evolving dark
energy EOS parameter $\omega_d$ and show a recent $\omega_d=-1$
crossing \cite{Bo}. As the dark energy model with a quintessence
field has an equation of state $\omega_d>-1$ and for the phantom
ghost field we always have $\omega_d<-1$, the phase transition
from quintessence to phantom era does not occur in models with
only one scalar field, either phantom or quintessence
\cite{divide}. To study such a transition, we need models which
are at least composed of two scalar fields \cite{linde}, known as
hybrid models.  One of these models is the quintom model which
assumes that the cosmological fluid, besides the matter and
radiation, is composed of a quintessence and a phantom scalar
fields \cite{quin}.  In \cite{kuo}, a phase-space analysis of a
spatially flat Friedman-Robertson-Walker universe containing a
barotropic fluid and phantom-scalar fields with exponential
potentials has been presented. This model has late time phantom
attractor solution. In \cite{quintpot} the same calculation has
been done by introducing an interaction term between the phantom
and the quintessence fields. The quantum stability of quintom
models is also an important problem whose some aspects have been
discussed in \cite{car}.

Recently a new view of quintom model, known as hessence model, has
been introduced. In this model,  the dark energy is described by a
single field with an internal degree of freedom rather than two
independent real scalar fields.  Hessence model can avoid the
difficulty of the Q-ball formation which gives trouble to the
spintessence model \cite{hessence}. The evolution of $\omega$ in
this model has been studied in \cite{cai} and \cite{our}, via
phase-space analysis. Although this model allows $\omega$ to cross
$-1$, but it avoids the late time singularity or the "big rip".

In the recent paper \cite{moh}, we have investigated the necessary
conditions for occurrence of the $\omega=-1$ crossing in the
quintom model ( which also implies the $\omega_d=-1$ crossing ),
in special the transition from quintessence to phantom phase. This
transition has been checked for two specific quintom potentials,
the power-law and exponential potentials, and has been shown that
the $\omega =-1$ crossing really exists in these examples. It has
been assumed that the fields have slowly varying behavior, i.e.
the problem has been solved in slow-roll (SR) approximation.

In this paper, we are going to prove that this phase transition
occurs for an arbitrary quintom potential, as long as we consider
the SR approximation. We think that this is an important result
which is in agreement with the present data.

The scheme of the paper is as follows. In section 2, we briefly
review the main results of \cite{moh} and write down the necessary
conditions needed for occurrence of the quintessence to phantom
phase transition in the quintom model. In section 3, by solving
the Friedman equations for arbitrary potential around the
transition point and in the SR approximation, it is shown that the
desired transition occurs for the general quintom model. The
general form of SR conditions, needed for consistency of the
equations, is stated in terms of the time derivatives of quintom
fields. In section 4, we obtain the explicit perturbative solution
of quintom fields for arbitrary potential, from which the SR
approximation can be expressed in terms of the derivatives of the
potential, which is more physical. This is discussed in appendix
A. We check our perturbative results with the exact expressions
known for special cases.

We use the units $\hbar=c=G=1$ throughout the paper.

\section{ Transition conditions in quintom model}
Consider a spatially flat Friedman-Robertson-Walker universe with
scale factor $a(t)$, filled with (dark) matter and quintom dark
energy. The evolution equation of matter density $\rho_m$ is
\begin{equation}\label{1}
\dot{\rho_m}+3H\gamma_m\rho_m=0,
\end{equation}
in which $\gamma_m=1+\omega_m$. $\omega_m$ is the equation of
state parameter of matter field defined by $\omega_m=P_m/\rho_m$,
where $P_m$ is the pressure of matter field.
$H(t)=\dot{a}(t)/a(t)$ is the Hubble parameter and "dot"
represents the time derivative. The quintom dark energy consists
of a normal scalar field $\phi$, i.e. the quintessence field, and
a negative kinetic energy scalar field $\sigma$, called the
phantom field. The energy density $\rho_D$ and pressure $P_D$ of
the homogenous quintom dark energy are \cite{quin,kuo}
\begin{eqnarray}\label{2}
\rho_D&=&\frac{1}{2}\dot{\phi}^2-\frac{1}{2}\dot{\sigma}^2+V(\phi,\sigma),\nonumber \\
P_D&=&\frac{1}{2}\dot{\phi}^2-\frac{1}{2}\dot{\sigma}^2-V(\phi,\sigma),
\end{eqnarray}
and their evolution equations are
\begin{eqnarray}\label{3}
&&\ddot{\phi}+3H\dot{\phi}+\frac{\partial V(\phi,\sigma)}{\partial \phi}=0,\nonumber \\
&&\ddot{\sigma}+3H\dot{\sigma}-\frac{\partial
V(\phi,\sigma)}{\partial \sigma}=0.
\end{eqnarray}
The Friedman equations, obtained from Einstein equations, are
\begin{eqnarray}\label{4}
H^2&=&\frac{8\pi}{3}(\rho_D+\rho_m)\nonumber \\
&=&\frac{4\pi}{3}[\dot{\phi}^2-\dot{\sigma}^2+2V(\phi,\sigma)+2\rho_m],
\end{eqnarray}
and
\begin{eqnarray}\label{5}
\dot{H}&=&-4\pi(\rho_D+\rho_m+P_D+P_m)\nonumber \\
&=&-4\pi(\dot{\phi}^2-\dot{\sigma}^2+\gamma_m\rho_m).
\end{eqnarray}
Note that eqs.(\ref{3}), (\ref{4}) and, (\ref{5}) are not
independent. The equation of state parameter
$\omega=(P_D+P_m)/(\rho_D+\rho_m)$ can be expressed in terms of
Hubble parameter as
\begin{equation}\label{1999}
\omega=-1-\frac{2}{3}\frac{\dot{H}}{H^2}.
\end{equation}
For $\dot{H}<0$, the system is in the quintessence phase
$\omega>-1$, and when $\dot{H}>0$, it is in the phantom phase with
$\omega<-1$. So crossing the $\omega=-1$ line is, in principle,
possible in quintom model.

If we are interested in situation in which the quintessence to
phantom phase occurs in some instant of time $t=t_0$, then $H(t)$
must have a local minimum at that time, i.e. $\dot{H}(t_0)=0$. So
at $t<t_0$, $\dot{H}<0$ and $\omega>-1$ and at $t>t_0$,
$\dot{H}>0$ with $\omega<-1$. If we restrict ourselves to
$t-t_0<<h_0^{-1}$, where $h_0=H(t_0)$ and $h_0^{-1}$ is of order
of the age of the universe, $H(t)$ can be taken as
\begin{equation}\label{7}
H(t)=h_0+h_1(t-t_0)^{\alpha}+O\left((t-t_0)^{\alpha+1}\right).
\end{equation}
$\alpha\geq 2$ is the order of the first non-vanishing derivative
of $H(t)$ at $t=t_0$ and $h_1= \frac{1}{\alpha
!}H^{(\alpha)}(t_0)$. $H^{(n)}(t_0)$ is the  $n$-th derivative of
$H(t)$  at $t=t_0$. The desired phase transition occurs when
$\alpha$ is an even positive integer and $h_1>0$.

For arbitrary quintom potential $V(\phi,\sigma)$, we want to study
if this situation ( even $\alpha$ and positive $h_1$ ) exists or
not. As usual, we restrict ourselves to SR approximation in which
the first terms of equations (\ref{3}) are negligible:
 \bea\label{8}
&&\ddot{\phi}<<H\dot{\phi},\nonumber \\
&&\ddot{\sigma}<<H\dot{\sigma}.
  \eea
\section{Seeking for transition solution}
Following \cite{moh}, we are going to find any solution to
eqs.(\ref{3})-(\ref{5}), when $H(t)$ given by eq.(\ref{7}) and the
fields vary slowly, i.e. eq.(\ref{8}), with the desired property,
that is $h_1>0$ and $\alpha$ = even.

Let us begin with eq.(\ref{5}) and expand it near $t_0\equiv 0$.
One finds
 \be\label{9}
 \alpha h_1t^{\alpha -1}+\cdots = -4\pi
 [\beta(0)+\dot{\beta}(0)t+\cdots ],
 \ee
in which
 \be\label{10}
 \beta(t)=\dot{\phi}^2-\dot{\sigma}^2+\gamma_m\rho_m.
 \ee
In terms of the dimensionless variables
$\dot{\Phi}=\dot{\phi}/h_0$, $\dot{\Sigma}=\dot{\sigma}/h_0$,
$R_m=\rho_m/h_0^2$, $\tau=h_0t$, and $H_1=h_1/h_0^{\alpha +1}$,
eq.(\ref{9}) can be written as
 \be\label{11}
 \alpha H_1\tau^{\alpha -1}+\cdots = -4\pi
 [B(0)+\frac{dB}{d\tau}(0)\tau+\cdots ],
 \ee
for $\tau<<1$ ( $t<<h_0^{-1}$ ). $B(t)$ is defined through
 \be\label{12}
 B(t)=\dot{\Phi}^2-\dot{\Sigma}^2+\gamma_mR_m.
 \ee
We will continue our study with eq.(\ref{9}), reminding that
shifting to dimensionless variables is always possible.

The matter density $\rho_m(t)$ can be found by solving
eq.(\ref{1}) with $H(t)$ given by eq.(\ref{7}). The result is
 \begin{equation}\label{13}
\rho_m(t)=\rho_m(0)e^{-3\gamma_m(h_0t+{h_1\over
{\alpha+1}}t^{\alpha+1})}.
\end{equation}
For $\alpha\geq 2$, the first equation obtained from eq.(\ref{9})
is
  \be\label{14}
 \beta(0)=\dot{\phi}^2(0)-\dot{\sigma}^2(0)+\gamma_m\rho_m(0)=0.
 \ee
For second relation, we first note that
  \be\label{15}
 \dot{\beta}(0)=2(\dot{\phi}\ddot{\phi}-\dot{\sigma}\ddot{\sigma})_0-3h_0\gamma_m^2\rho_m(0).
 \ee
But in SR approximation (\ref{8}), we have
  \be\label{16}
  -(H\dot{\phi}-\ddot{\phi})\dot{\phi}+(H\dot{\sigma}-\ddot{\sigma})\dot{\sigma}\simeq H(
  \dot{\sigma}^2-\dot{\phi}^2).
  \ee
So
  \be\label{17}
 (\dot{\phi}\ddot{\phi}-\dot{\sigma}\ddot{\sigma})_0
  << h_0(\dot{\sigma}^2-\dot{\phi}^2)_0=h_0\gamma_m\rho_m(0),
  \ee
where in the last equality we use eq.(\ref{14}). Therefore,
excluding special $\gamma_m<<1$ cases, one finds
  \be\label{18}
   -4\pi\dot{\beta}(0)=12\pi h_0\gamma_m^2\rho_m(0),
 \ee
which is a non-zero positive quantity. This shows that for all
quintom potentials, one has
  \bea\label{19}
   &&\alpha=2\nonumber\\ && h_1=6\pi h_0\gamma_m^2\rho_m(0)>0,
   \eea
which proves the $\omega =-1$ crossing for all quintom models in
SR approximation, {\it as long as} the remaining relations which
are obtained from eq.(\ref{4}), are satisfied, up to the lowest
order, consistently.

The important observation is that in all quintom models in SR
approximation, the rate of the phase transition ($h_1$) only
depends on the (dark) matter energy density $\rho_m(0)$, which
itself is determined by the kinetic part of quintom energy density
(eq.(\ref{14})).

It is worth noting that if there is no dark matter field, the
eq.(\ref{18}), and therefore eq.(\ref{19}), can not be used
directly. In this case, we can not neglect the first two-terms of
eq.(\ref{15}) and therefore $h_1$ becomes
 \be\label{19-1}
 h_1=-4\pi(\dot{\phi}\ddot{\phi}-\dot{\sigma}\ddot{\sigma})_0.
 \ee
The above equation shows that the $\omega =-1$ crossing is in
principle possible in dark-matter free quintom models.

To prove the consistency of the remaining relations, we expand
eq.(\ref{4}) near $t=0$, which results in
  \be\label{20}
  h_0^2+2h_0h_1t^\alpha+\cdots=\frac{4\pi}{3}[\delta(0)+\dot{\delta}(0)t
  +\frac{1}{2}\ddot{\delta}(0)t^2+\cdots],
  \ee
in which
  \be\label{21}
  \delta(t)= \dot{\phi}^2-\dot{\sigma}^2+2V(\phi,\sigma)+2\rho_m.
  \ee
The first two relations obtained from eq.(\ref{20}) are
  \be\label{22}
  \frac{8\pi}{3}\left[
  V(0)+(1-\frac{\gamma_m}{2})\rho_m(0)\right]=h_0^2,
  \ee
which determines $h_0$ in terms of $V(0)$ and $\rho_m(0)$, and
$\dot{\delta}(0)=0$, which can be written as
  \be\label{23}
  (\dot{\phi}\ddot{\phi}-\dot{\sigma}\ddot{\sigma}-3h_0\gamma_m\rho_m
   + \dot{V})_{t=0}=0.
 \ee
But
  \be\label{24}
  \dot{V}=\frac{\partial V}{\partial \phi}\dot{\phi}+
  \frac{\partial V}{\partial \sigma}\dot{\sigma} =
  3H(\dot{\sigma}^2-\dot{\phi}^2),
  \ee
in which the equation of motion (\ref{3}) in SR approximation has
been used. So eq.(\ref{23}) can be written as
  \be\label{25}
  \left(\dot{\phi}(\ddot{\phi}-3H\dot{\phi})-\dot{\sigma}(\ddot{\sigma}
  -3H\dot{\sigma})-3h_0\gamma_m\rho_m\right)_{t=0}=0,
  \ee
which is reduced to eq.(\ref{14}) in SR approximation.

The last equation that must be checked in the lowest order
approximation, is one obtained from $t^2$-term in the
right-hand-side of eq.(\ref{20}). Using eqs.(\ref{13}) and
(\ref{24}) and $\dot{H}(0)=0$, it becomes
  \bea\label{26}
  &&\frac{4\pi}{3}[\ddot{\phi}^2-6h_0\dot{\phi}\ddot{\phi}
  -\ddot{\sigma}^2+6h_0\dot{\sigma}\ddot{\sigma}+
  \dot{\phi}\ \dddot{\phi}-\dot{\sigma}\ \dddot{\sigma}+
  9h_0^2\gamma_m^2\rho_m]_{t=0}\nonumber \\&\simeq&
  \frac{4\pi}{3}[\dot{\phi}(\dddot{\phi}-6h_0\ddot{\phi})
  +\dot{\sigma}(-\dddot{\sigma}+6h_0\ddot{\sigma})+
    9h_0^2\gamma_m^2\rho_m]_{t=0},
    \eea
where the SR approximation has been used in the second equality.
But the natural extension of SR approximation (\ref{8}) to higher
order terms is
  \bea\label{27}
  \frac{\d^n\phi}{\d t^n}&<<&h_0 \frac{\d^{n-1}\phi}{\d t^{n-1}},
  \nonumber \\
   \frac{\d^n\sigma}{\d t^n}&<<&h_0 \frac{\d^{n-1}\sigma}{\d
   t^{n-1}}.
   \eea
So the eq.(\ref{26}) reduces to
  \be\label{28}
  4\pi h_0[2(\dot{\sigma}\ddot{\sigma}-\dot{\phi}\ddot{\phi})+3h_0\gamma_m^2\rho_m]_{t=0}
  =-4\pi h_0\dot{\beta}(0)=2h_0h_1,
 \ee
where eqs.(\ref{15}) and (\ref{9}), with $\alpha =2$, have been
used. Now this is exactly equal to the second term in the
left-hand-side of eq.(\ref{20}) for $\alpha=2$.

Let us briefly express what we have done. We have shown that the
eqs.(\ref{3})-(\ref{5}) have a transition solution when $H(t)$
behaves as (\ref{7}) with $\alpha=2$. It has been shown that
eqs.(\ref{4}) and (\ref{5}) result three independent relations:
Eq.(\ref{14}), which relates $\rho_m(0)$ to the kinetic energy of
quintom fields, and eqs.(\ref{19}) and (\ref{22}) which determine
the Hubble parameters $h_0$ and $h_1$ in terms of matter field and
potential. For $t<<h_0^{-1}$ and in SR approximation, it has been
shown that all remaining relations are consistent with the
mentioned three ones. This completes the proof of $\omega=-1$
crossing of {\it all} quintom models in SR approximation.$\square$

It is worth noting that the validity of eq.(\ref{27}) can be
verified by following argument. Consider the equation of motion
(\ref{3}) near $t=0$. Differentiating these equations with respect
to $t$, and using $\dot{H}(0)=0$, one finds
  \bea\label{29}
  &&\dddot{\phi}+3H\ddot{\phi}+\frac{\d }{\d t}\frac{\partial
  V}{\partial \phi}=0,\nonumber \\
  &&\dddot{\sigma}+3H\ddot{\sigma}-\frac{\d }{\d t}\frac{\partial
  V}{\partial \sigma}=0.
  \eea
Now ignoring the $\ddot{\phi}$ ( $\ddot{\sigma}$ ) term in
eq.(\ref{3}) in comparison with $H\dot{\phi}$ ( $H\dot{\sigma}$ ),
is equivalent to ignoring the $\dddot{\phi}$ ( $\dddot{\sigma}$ )
term in eq.(\ref{29}) in comparison with $H\ddot{\phi}$ (
$H\ddot{\sigma}$ ). Repeating this differentiation procedure
justifies eq.(\ref{27}) as the general SR conditions.

\section{Solution of equations of motion}
In special cases, like ones considered in \cite{moh}, one can, in
principle, solve the equations of motion (\ref{3}) in SR
approximation and studies their physical behaviors. For arbitrary
potential, as expected, we can not solve these equations without
knowing the functional form of the potential, but instead we try
to find a perturbative solution, from which one can deduce some
physics behind the problem. Here we focus on the equation of
motion of the quintessence field. The solution of phantom field
can be simply obtained from final relations by replacing the
derivatives from $\phi$ to $\sigma$ and transforming $V\rightarrow
-V$.

Consider the equation of motion of quintessence field in SR
approximation $\ddot{\phi}<<H\dot{\phi}$ with $H=h_0+h_1t^2$:
  \be\label{30}
  3(h_0+h_1t^2)\dot{\phi}=-\frac{\partial V(\phi,\sigma)}{\partial
  \phi}.
  \ee
Using the Lerchphi function $\Phi(z,a,b)$ defined as
  \be\label{31}
  \Phi(z,a,b)=\sum_{n=0}^\infty\frac{z^n}{(b+n)^a},
  \ee
it can be easily shown
  \be\label{32}
  \frac{\d}{\d t}\left[
  \frac{t\Phi\left(-\frac{h_1t^2}{h_0},1,\frac{1}{2}\right)}{2h_0}\right]
  =\frac{1}{h_0+h_1t^2}.
  \ee
Eq.(\ref{30}) then leads to
  \be\label{33}
  \frac{t}{2h_0}\Phi\left(-\frac{h_1t^2}{h_0},1,\frac{1}{2}\right) =
  3G(\phi)+\lambda,
  \ee
where
  \be\label{34}
  G(\phi)=-\int\frac{\d \phi}{\partial V/\partial
  \phi},
  \ee
and $\lambda$ is the constant of integration. Note that in this
stage, we assume that $\partial V/\partial\phi$ does not depend on
$\sigma$. Expanding both sides of eq.(\ref{33}) around $t=0$ up to
order $t^3$, results in
  \be\label{35}
  \frac{t}{h_0}-\frac{h_1}{3h_0^2}t^3+O(t^5)= 3[G(0) +\dot{G}(0)t
  +\frac{1}{2!}\ddot{G}(0)t^2+ \frac{1}{3!}\dddot{G}(0)t^3 +
  O(t^4)]+\lambda ,
  \ee
which leads to
  \bea\label{36}
  3G(0)+\lambda &=& 0,\nonumber\\
  3\dot{G}(0)&=&\frac{1}{h_0},\nonumber\\
  \ddot{G}(0)&=&0, \\
  \frac{1}{2}\dddot{G}(0)&=&-\frac{h_1}{3h_0^2},\nonumber\\
  \vdots &&\nonumber
  \eea
Using
  \bea\label{37}
  \dot{G}&=&G'\dot{\phi}=-\frac{1}{A}\dot{\phi},\nonumber\\
  \ddot{G}&=&\frac{B}{A^2}\dot{\phi}^2-\frac{1}{A}\ddot{\phi},\\
  \dddot{G}&=&\frac{AC-2B^2}{A^3}\dot{\phi}^3+\frac{3B}{A^2}\dot{\phi}
  \ddot{\phi}-\frac{1}{A}\dddot{\phi},\nonumber
  \eea
where
 \be\label{38}
 A=\left(\frac{\partial V}{\partial \phi}\right)_0,  \ \ \
 B=\left(\frac{\partial^2 V}{\partial \phi^2}\right)_0, \ \ \
 C=\left(\frac{\partial^3 V}{\partial \phi^3}\right)_0,
  \ee
one can obtain $\dot{\phi}(0)$, $\ddot{\phi}(0)$, and
$\dddot{\phi}(0)$ from eq.(\ref{36}), from which $\phi(t)$ is
obtained up to order $t^3$ as following
  \be\label{39}
  \phi(t)=\phi(0)-\frac{A}{3h_0}t+ \frac{AB}{18h_0^2}t^2
  -\frac{A}{18h_0^2}\left(\frac{B^2+AC}{9h_0}-2h_1\right)t^3
  +O(t^4).
  \ee
This is our desired expression.

As an example, we consider the exponential potential
  \be\label{40}
  V=v_1e^{\lambda_1\phi}+v_2e^{\lambda_2\sigma}.
  \ee
Eq.(\ref{39}) then results in:
  \be\label{41}
  \phi(t)=\phi_0-\frac{v_1\lambda_1e^{\lambda_1\phi_0}}{3h_0}t+
   \frac{v_1^2\lambda_1^3e^{2\lambda_1\phi_0}}{18h_0^2}t^2
  -\frac{v_1\lambda_1e^{\lambda_1\phi_0}}{18h_0^2}\left(\frac{2v_1^2\lambda_1^4
  e^{2\lambda_1\phi_0}}{9h_0}-2h_1\right)t^3
  +O(t^4).
  \ee
But it can be easily shown that the above expression is same as
one obtained from expanding the exact expression: \cite{moh}
  \be\label{42}
  \phi(t)={1\over {\lambda_1}}\ln\left\{{6h_0
  \over{v_1\lambda_1^2\left[ t\Phi({-h_1t^2\over
  h_0},1,{1\over 2})+2c_1h_0\right]}}\right\},
  \ee
in which
  \be\label{43}
  c_1=\frac{3}{v_1\lambda_1^2}e^{-\lambda_1\phi_0}.
  \ee

At the end, let us discuss a possible ambiguity. It may be argued
that as we take $H(t)$ up to order $t^2$ in eq.(\ref{30}), keeping
the $t^3$-terms for $\phi(t)$ is not correct. To answer this
question, we keep the $t^3$-term in $H(t)$:
  \be\label{49}
  H(t)=h_0+h_1t^2+h_2t^3+\cdots
  \ee
In SR approximation, we have
  \be\label{50}
  \int\frac{\d t}{H(t)}=3G(\phi)+\lambda.
  \ee
Using (\ref{49}), the left-hand-side of eq.(\ref{50}), up to order
$t^3$, is
  \be\label{51}
   \frac{t}{h_0}-\frac{1}{3}\frac{h_1}{h_0^2}t^3-
   \frac{1}{4}\frac{h_2}{h_0^2}t^4+\cdots
   \ee
So the coefficient $h_2$ only contributes to $t^4$-term and
therefore the expansion (\ref{39}) is correct.

 {\bf Acknowledgement:} We would
like to thank the "center of excellence in structure of matter" of
the Department of Physics for partial financial support.\\ \\

\setcounter{equation}{0}
\renewcommand{\theequation}{A.\arabic{equation}}
{\bf\large Appendix A: general SR condition for potential}\\

 In this appendix, we want to obtain the analogous of SR conditions (\ref{27}) for
quintom potential, which is probably more physical. This can be
done by using the explicit solution (\ref{39}). For simplicity, we
assume that the parameters $A$, $B$, and $C$, defined in
eq.(\ref{38}), are positive real numbers. Using (\ref{39}), we
first note that the condition $|\ddot{\phi}|<<h_0|\dot{\phi}|$
simply results in
  \be\label{44}
  B=\left(\frac{\partial^2 V}{\partial \phi^2}\right)_0<<h_0^2.
  \ee
This constraint can be also obtained by noting that in SR
approximation, we have $3h_0\ddot{\phi}(0)+\d/\d t(\partial
 V/\partial\phi)|_0=0$ ( see eq.(\ref{29}) ), which results
 $\ddot{\phi}(0)=-(1/3h_0)(\partial^2V/\partial
 \phi^2)\dot{\phi}|_0$. Then $|\ddot{\phi}|<<h_0|\dot{\phi}|$
 reduces to eq.(\ref{44}).

The next conditions can be obtained by nothing that the
consistency of $t^2$-term of eq.(\ref{20}) can be achieved if the
$(\dot{\phi}\ \dddot{\phi})_0$ ( and $(\dot{\sigma}\
\dddot{\sigma})_0$ ) term in eq.(\ref{26}) is negligible in
comparison with $h_0^2\rho_m(0)$ term. But eq.(\ref{14}) shows
that $\rho_m(0)\sim\dot{\phi}^2(0)$, so the equations are
consistent if
  \be\label{45}
  \frac{1}{h_0^2}(\dot{\phi}\ \dddot{\phi})_0<< \dot{\phi}^2(0).
  \ee
Using eq.(\ref{39}), eq.(\ref{45}) reduces to
  \bea\label{46}
   \frac{1}{h_0^2}\frac{-A}{3h_0}\frac{-A}{3h_0^2}
   \left(\frac{B^2+AC}{9h_0}-2h_1\right)&<<& \frac{A^2}{9h_0^2}
   \ \ \Rightarrow   \nonumber\\
   \frac{1}{9}\left(\frac{B}{h_0^2}\right)^2- 2\frac{h_1}{h_0^3}
   +\frac{1}{9}\frac{A}{h_0^2}\frac{C}{h_0^2}&<<&1.
   \eea
Now  eq.(\ref{19}) implies
  \be\label{47}
  \frac{h_1}{h_0^3}\sim\frac{\rho_m(0)}{h_0^2}\sim
  \frac{\dot{\phi}^2(0)}{h_0^2}\sim \frac{A^2}{h_0^4},
  \ee
where eqs.(\ref{14}) and (\ref{39}) have been used. So
eq.(\ref{46}) is satisfied if, besides condition (\ref{44}), one
has $A<<h_0^2$ and $C<<h_0^2$. The generalization is obvious: the
general SR conditions, in the language of quintom potential, are:
  \bea\label{48}
  && \left(\frac{\partial^n V}{\partial \phi^n}\right)_0<<h_0^2,
  \nonumber\\
   && \left(\frac{\partial^n V}{\partial \sigma^n}\right)_0<<h_0^2,
   \eea
which indicate the slowly varying potential.

\end{document}